# Cloud Computing: Exploring the scope

Indian Institute of Information Technology- Allahabad
Abhinav Pandey, Akash Pandey, Ankit Tandon, Brajesh Kr. Maurya, Upendra Kushwaha
Dr. Madhvendra Mishra, Vijayshree Tiwari

**Abstract**

*Cloud computing refers to a paradigm shift to overall IT solutions while raising the accessibility, scalability and effectiveness through its enabling technologies. However, migrated cloud platforms and services cost benefits as well as performances are neither clear nor summarized. Globalization and the recessionary economic times have not only raised the bar of a better IT delivery models but also have given access to technology enabled services via internet. Cloud computing has vast potential in terms of lean Retail methodologies that can minimize the operational cost by using the third party based IT capabilities, as a service. It will not only increase the ROI but will also help in lowering the total cost of ownership. In this paper we have tried to compare the cloud computing cost benefits with the actual premise cost which an organization incurs normally. However, in spite of the cost benefits, many IT professional believe that the latest model i.e. "cloud computing" has risks and security concerns. This report demonstrates how to answer the following questions: (1) Idea behind cloud computing. (2) Monetary cost benefits of using cloud with respect to traditional premise computing. (3) What are the various security issues? We have tried to find out the cost benefit by comparing the Microsoft Azure cloud cost with the prevalent premise cost.*

## 1. Introduction

Cloud computing is an emerging IT service model paving its way into the business world. Understanding of cloud differs from people to people. A technological view is quite common while another perspective support the role of communication enabler. In layman words, "Cloud computing is about moving services, computation and/or data off-site to an internal or external, location-transparent, centralized facility or contractor. [1]" It provides the functionality of deploying scalable IT resources over the Internet. Cloud provides application and services without taking into account the actual cost of the infrastructure and the software. It helps the business to avoid capital expenditure and operational expenditure Y-o-Y. In cloud model the user "pay as you go" [2] to the services provided. Thus, it helps in seizing a lot of unnecessary expenses for the organization. This money can later be used to increase the company's core competency.

Cloud computing is a way different from traditional computing. In traditional approach the organization need to set up large servers, storage devices and lot of other equipments to support the business. This not only need large capital expenditures but also requires lot of effort in terms of maintenance and up gradation of technology. Uninterrupted power supply, cooling mechanism, expert technical team, big server rooms were undoubtedly the other requirements. Now due to this model i.e. cloud computing all such requirements are eliminated [3]. The users just need to pay as they use. Three major categories of services are provided by cloud computing system [4].

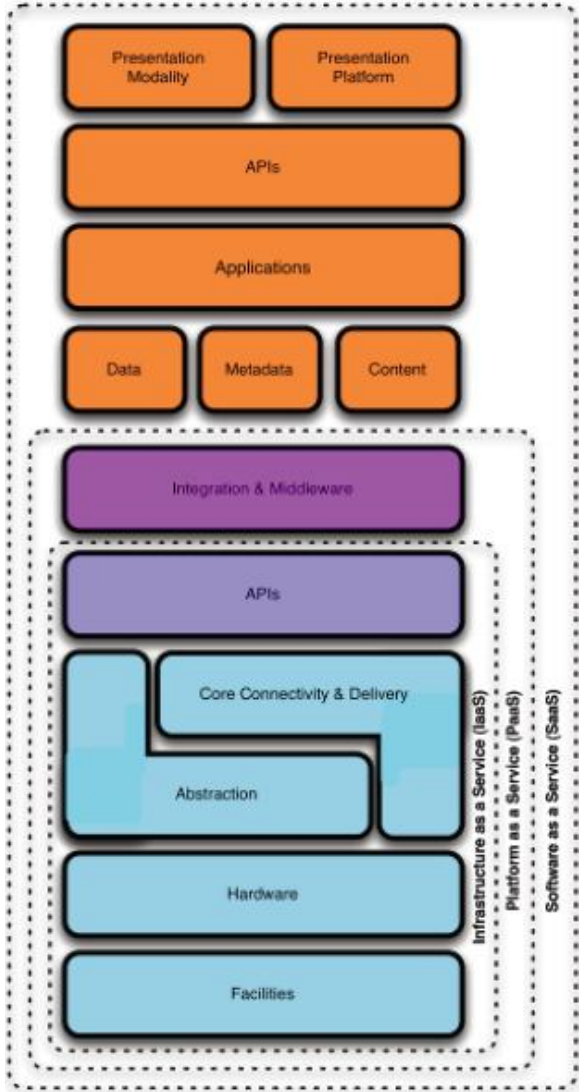

**Figure 1 Cloud reference model**

**Infrastructure as a Service (IaaS)**

Like, Amazon web service provides virtual servers with unique IP addresses and block of storage on demand. Customers benefit from an application program interface (API) and customer can pay for exactly the amount of service they use, like for electricity and water. This model is also called utility computing.

**Platform as a Service (PaaS)**

It is a set of software and development tool hosted on provider servers. Developers can create application using the provider APIs. Google Apps Engine is one of the most famous platforms as a service provider [5].

**Software as a Service (SaaS)**

In this case, the provider allows the customer only to use its application. The software interacts with the user with the user interface. This application can be anything from web based email like yahoo, hotmail, Gmail to applications like twitter, facebook and orkut.

Though cloud computing has stormed the whole business world, but yet the cost benefits are not perfectly clear. It is not certain that this model will help every business in the same way. It is also unambiguous that this latest model is really helpful for the organizations. In this paper we try to figure out what are the cost benefits of moving to cloud by comparing the traditional approach with latest technology. We also try to find out the answers to the following questions.

- Is the cloud relevant for the retail sector?
- What are the computing rates and the associated cost when organizations move to cloud?
- Can cloud be used along with traditional approach to increase cost efficiency?

To answer these questions, we collected data from several organizations using traditional computing system and performed crash costing to compare it with the cloud computing cost as per the Amazon web services.

**2. Related Work**

One of the most challenging issue that hinders the growth of cloud computing as the future technology is the lack of a unanimous definition. Two major institutions have significantly contributed in clearing the fog, National Institute of standards and technology and cloud security alliance. They both agree to a definition of cloud that "it is a model for enabling convenient, on-demand network access to a shared pool of configurable computing resources (e.g. networks, servers, storage, and applications and services) that can be rapidly provisioned and released with minimal management effort or service provider interaction."[5]

Significant amount of study and researches have been done to analyze the cost effectiveness and business utility of cloud computing. [7] Provides an insight into the major drivers enabling the retailers for the adoption of cloud computing as an alternate of traditional computing. It further explores the relevance and priority of issues like vendor preference, business collaboration and cost benefit for the retailers.

In [8] monetary and performance comparison of cloud computing vs. grid computing was figured out. A comprehensive approach was adopted thereby collecting on spot server performance data and financial expenses incurred. In [9] researchers studied the case for organizations curious of expanding their IT infrastructure by utilizing the cloud applications and services. Their research try to explore the various strategies that might be help while trying to schedule the optimum utilization of computing resources i.e. both cloud and grid models. Analysis of the cloud computing business and security perspectives was carried out in [10] which describes cloud as a very promising technology capable of improving the business performance and may deliver a whole bunch of benefits but it's quite essential to carry out risk assessment and business impact analysis therefore appraising the enterprises possible threats of shifting to cloud. In [11] major drivers behind cloud computing adoption and it's three major business utilities speed, focus and finance were discussed. Cloud computing enables the commencing of a new project within specified time period, reason being cloud starts with a pre-build foundation. By the way of cloud an enterprise earn the ability to focus over its core competency and outsource the remaining business functions. In place of huge upfront investment, must for establishing, implementing, operating, maintaining, reviewing and upgrading the IT capabilities, cloud let you feel the burden of your present expenses only, therefore acting as a rescuer in crunch times. It marked out essential phases that an organization must follow while thinking of cloud

- Set up a performance realization metric which replicates the value generated as the use of the application grows.
- Pragmatic approach i.e. think big, act small. Set up a pilot project and translate its success into enterprise wide execution incorporating all essential milestones which quantify the detriment in terms of usage impact.
- An industry wide analysis to get the most beneficial services and application as per requirement of the organization.

**3. Indian Retail Industry:**

"Retailing is act of selling goods or commodities in small quantities directly to consumers." Indian retail sector is one among those industries riding on the bull these days. Indian retail sector has shown enormous potential in recent few years and has maintained an average growth rate of around 30% annually [12]. It's also termed out as most lucrative investment destination for the Foreign Direct investment (FDI) as per the AT Kearney 8[th] Annual global retail Development Index [13]. Major drivers behind this growth factor are:

- 60% of the Indian population is below the age of 30 [14].
- A major chunk of Indian population dwells in metropolitan cities where most of the retailers are located.
- Significant hike in disposable income of the middle class household therefore flourishing the "Mall Culture".
- Impact of globalization: international exposure, entry of various brands.

The composition of Indian retail sector is very interesting attributed to the fact that its major chunk comprised of unorganized sector quite contrary to some other major retail destination over world but this trend is watching a change nowadays. Indian retail industry is passing through a stage of transformation or flux where the huge capacity and vast potential of it has lured the major retail giants across the world. It is therefore posing a threat before the unorganized sector either to stay in competition by offering a value proposition to the customer or being wiped out of the scenario. Major retail players entering into the field are also feeling the heat of competition from the unorganized sector. New Customer attraction and retention are the critical success factors in such a throat cut competition. In such a scenario where customer experience plays a vital role it becomes quite essential for any retailer to provide customer most delight experience. This is field where retailers looks toward the IT and expect that IT can help them in creating a better shopping ambience and experience.

Some of the IT innovations have brought a revolutionary change not only in the back end operations but also offering a significant change in the customer delivery and customer support. Innovations introduced in the retail industry are multi spine and aimed towards better resource utilization and optimized logistics, delight user experience, and to match the pace of ever changing industry demands. These innovations encompass all activities ranging right from the supplier management to customer relationship management. Some of these innovations are:

- Radio frequency identification (RFID): Enable to store the customer related information like preferences, shopping behavior etc on a smart chip. It was also used for the product to reduce the time taken at the sales counter.
- E-catalog based selling: product is introduced to the customer with the help of self exploring kiosks.
- Automated tracking system: Retail giants like Wal-Mart uses GPS system along with RFID technology to locate products. It makes shopping an easy experience for the customer.
- Deployment of applications like SAP, ERP, CRM, supply chain optimization and master data repository.

Cloud computing is an emerging concept making its place into the industrial scenario. Major objective of a business remains to be generating profits thereby maximizing the value generated for the customer. It can be achieved through integrated and streamlined processes, enhanced productivity and offering a better solution to the customer. Concept of cloud has emerged as a savior for the retailers therefore restructuring the basic IT configuration which remained unaltered for last many years. Cloud computing offers business dimensions of flexibility, interoperability and fast turnaround time to the retailers therefore more retailers are inclined towards the adoption of cloud or planning to adopt it as shown in figure 1[15].

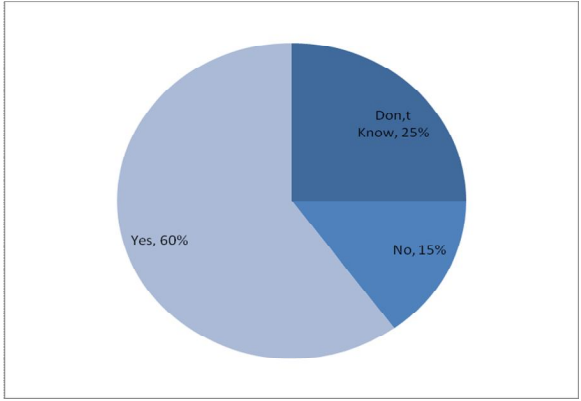
**Figure 2 Emergence of "Cloud" in retail sector**

## 4. Cloud Computing Potential:

There are lots of benefits that an organization can get by switching to cloud model. Some of them are:

**Cost containment**— The new model not only help to minimize the hardware and software cost, but it also help in dealing with the scalability issues. Moreover, an organization does not need to purchase the new infrastructure. The new model help in giving them metered service. In this model the user just need to give the amount for which he had used in the form of the service. The model also helps the enterprise to minimize the cost as it help to save the unused server space. This will also give them a chance to use the new technology without worrying about the extra IT cost. Before switching to the cloud service every organization must have a clear idea about the cost benefits, they will get, if they change their IT setup from their traditional on premise to cloud based service.

**Immediacy**— The cloud based service provide us a application in relatively lesser time rather than the traditional approach. Prior methods require months to integrate and then implement the resource. It has not only helped to lower down the cost but also can reduce the time delays.

**Availability**— The most essential part is that the cloud providers should have the latest hardware, software and bandwidth that is needed for their business. Their clients want high speed access, large storage space and applications to run in their own enterprise. The service provider should provide the infrastructure, platform such that there is effective load balancing and the servers do not get overloaded even at the time of high traffic. Though the availability of services is guaranteed from provider side yet the client should make an alternative arrangement to deal with the service interruption.

**Scalability**— Cloud provider offers more flexibility to increase the needed infrastructure and the services according to the need of the clients. Also, it will reduce the valuable time needed to offer a new service. Everything is done as soon as demand is created from the client side.

**Efficiency**— Since this model help to minimize the cost therefore, the organizations get an opportunity to utilize the savings in some innovation and research and development. This allows a business to use the benefited cost to improve their core competencies. It certainly helps to be more advantageous in longer run.

**Resiliency**— The service providers have made an arrangement by which they guarantee to provide the services even in case of natural disaster. They ensure that they are capable enough to sustain through any kind of unexpected scenario.

## 5. Drivers for Cloud Computing:

The major drivers which are compelling the growth of cloud in retail are very basic in nature, huge IT expenditure, Enhanced hardware efficiency and the ability to manage the uncertainty associated with future store and channel server capacity demand and a centralized data center [7]. While current traditional infrastructure require a huge capital expenditure as well as ongoing operational expenditure cloud computing presents a lucrative and viable solution. Cloud computing means better, faster and more relational interactions with customers, suppliers and partners as systems that can interoperate and are more agile in nature. Another most significant benefit offered through cloud is flexibility. As a retailer grows in its operation, its technological infrastructure and capabilities need to be redefined in order to support the expansion. Figure 2 shows the major business factors most significant according to retailers for propelling the growth of cloud[7].

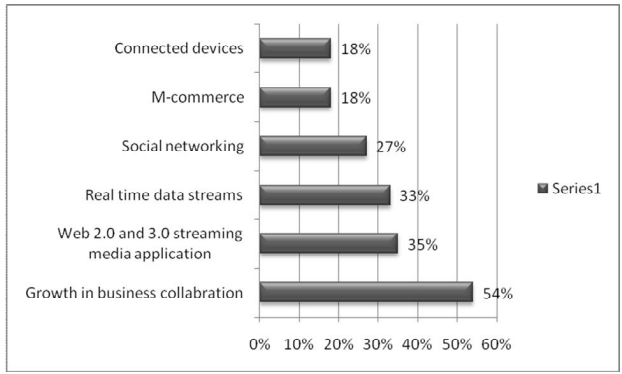

**Figure 3 Driver propelling the growth of cloud**

Some other vital factors playing a crucial role towards the adoption of cloud computing results from business movement of concentrating over their core competency while rest of the activities must be outsourced. Retailers today are trying to have a decreased IT spends over the deployment, development of IT applications, help desk support and human resources. An enhanced awareness towards the "Green technology Initiative", information security and disaster recovery standards also act as a motivating factor for the organizations to go for cloud. Figure 3 explains the reasons for moving to cloud computing [15].

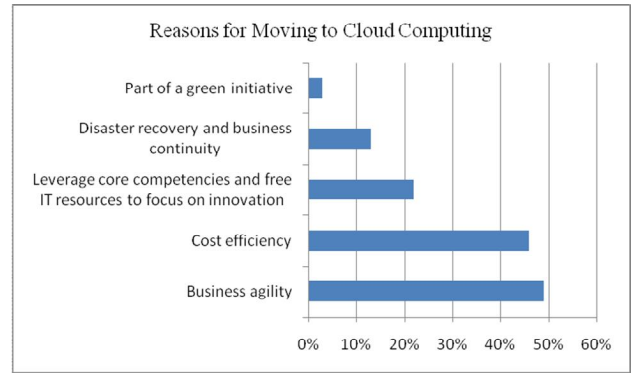

**Figure 4: Reasons for moving to cloud computing**

Apart of the major business factors, for a retailer, some technological issues are must to be taken into consideration which involves escalation in IT costs, improvement in hardware efficiency and capability for an organization to consider future channel server capacity demands and need of a centralized data center. In recent years, cloud computing has gone through a huge dip in its connectivity and operational cost prompting retailers towards it.

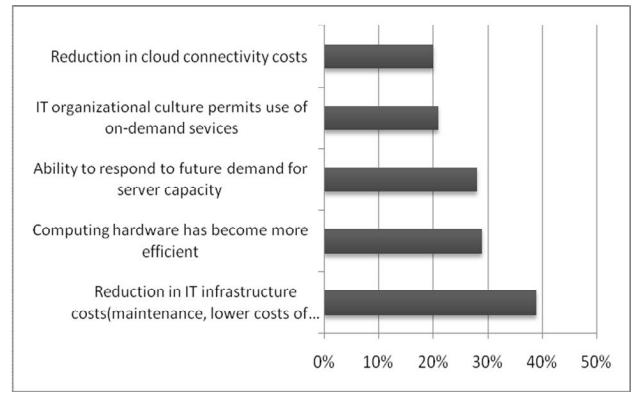

**Figure 5: Focus on cloud driven by IT costs and Hardware Needs**

Also, an increase efficiency of the hardware capability and server capacity is important. Figure 4 represents technological factors mainly IT costs and hardware needs pushing the growth of cloud [7].

# 6. Cost benefit analysis

|  | On-premises | | | | Cloud Based | | | |
|---|---|---|---|---|---|---|---|---|
|  | Unit | Qty | Per Unit (in '000) | Total (in '000) | Unit | Qty | Per Unit | Total (in '000) |
| **CAPEX** | | | | 163610 | | | | 480 |
| Server | No. | 100 | 225 | 22500 | No. | 0 | 0 | 0 |
| Storage | TB | 50 | 155 | 7750 | TB | 0 | 0 | 0 |
| Storage (Backup) | TB | 250 | 45 | 11250 | TB | 0 | 0 | 0 |
| Software (OS) | No. | 100 | 180 | 18000 | No. | 0 | 0 | 0 |
| Software (DB) | No. | 100 | 1000 | 100000 | No. | 0 | 0 | 0 |
| Software (Anti Virus) | No. | 100 | 9 | 900 | No. | 0 | 0 | 0 |
| Networking | No. | 50 | 45 | 2250 | No. | 0 | 0 | 0 |
| Man Power | Rs/res/pa | 10 | 8 | 960 | Rs/res/pa | 5 | 8000 | 480 |
|  | | | | | | | | |
| **OPEX (Annual)** | | | | 37598 | | | | 34643.9 |
| Computing Power | | 0 | 0 | 0 | Rs/hr | 6307200 | 5.40 | 34058.9 |
| Bandwidth | Rs/pa | 3 | 1550 | 4650 | Rs/pa | 15 | 250 | 45 |
| Staff Salary | Res/year | 15 | 15 | 2700 | Res/year | 3 | 15 | 540 |
| Infra. Maintenance | 15% of TC | 0 | - | 6563 | 10% of TC | 0 | 0 | 0 |
| Software Maintenance | 15% of TC | 0 | - | 17835 | 10% of TC | 0 | 0 | 0 |
| Electricity | Rs/pa | 0 | 0.005 | 3000 | Rs/pa | 0 | 0 | 0 |
| Rent for Real Estate | Rs/sft/pa | 1500 | .075 | 1350 | Rs/sft/pa | 0 | 0 | 0 |
| Other Maintenance | Rs/year | 0 | - | 1500 | Rs/year | 0 | 0 | 0 |
| **Total** | | | | 201208 | | | | 35123.9 |
| **Savings** | 166084.1 | | | | | | | |

1INR=0.0220458USD (Source: http://www.xe.com/ucc/) 14[th] May, 2010, 7:00 PM (IST)

In Indian context, several assumptions are made for this analysis. It gives us the idea of monetary benefits of the cloud based model with respect to the normal on premise scenario. It clearly depicts that there is a drastic change in the overall IT capital expenditure. The on premise capital expenses are very high as compared to outsourced services. The main components include the cost of servers, storage and software with licenses. These include 90% of the overall capital expenditure. Whereas operational expenditure is also reduces Y-o-Y [20]. It is difficult for every CFO to reduce IT operation cost. Third party IT hardware and software services are the final solution. The cloud service provider offers the storage space, IT hardware and software in accordance with the need of the user. The users just have to pay the used amount. It's same as paying bill of water or electricity.

## 7. ROI framework

|  | On-premises | | | | Cloud Based | | | Cumulative Savings | Y-o-Y savings (%) |
|---|---|---|---|---|---|---|---|---|---|
|  | CAPEX (initial) | OPEX (annual) | Cumulative Cost | | CAPEX (initial) | OPEX (annual) | Cumulative Cost | | |
| Year-1 | 163610 | 37598 | 201208 | | 480 | 34643.9 | 35123.9 | 166084.1 | **82.54** |
| Year-2 | - | 37598 | 238806 | | - | 34643.9 | 69767.8 | 169038.2 | **70.78** |
| Year-3 | - | 37598 | 276404 | | - | 34643.9 | 104411.7 | 171992.3 | **62.22** |
| Year-4 | - | 37598 | 314002 | | - | 34643.9 | 139055.6 | 174946.4 | **55.71** |
| Year-5 | - | 37598 | 351600 | | - | 34643.9 | 173699.5 | 177900.5 | **50.59** |

Y-o-Y saving in cloud model is given below. Initially, we have calculated the Return on investment for 4 years

## 8. Security issues

Cloud computing no doubting stands as one of the most promising and enticing areas in technological advancement as far as its ability of flexibility and cost effectiveness is concern. Despite of the interest it has generated and its ability to affect business efficiency and performance there exist some significant and persistent concerns over the authenticity and reliability of cloud computing. Gartner contributed a significant contribution towards identification, prioritization, and suggesting some remediation for the same [17], [19].

The major cloud security risks identified by Gartner are [19]:

- Privileged user access
- Regulatory compliance
- Data segregation
- Data recovery
- Investigative Support
- Long term Viability

## 9. Conclusion

As demonstrated in this report, the top two benefits of using cloud computing in retail are Y-o-Y savings through lower cost of ownership (48%) and return on investment (62%) [7]. Also, it helps to build an insight for retailers to consider third party cloud services as an alternate to conventional use of owned IT hardware and software capabilities. Speculation in the cost benefit analysis of both ownership (On-premises) and subscribership (cloud based) scenario motivates the retailers to explore the cloud computing capabilities to minimize the cost of operation, leverage core competencies and enhances the business agility [15]. The findings of this report also showcases the major technology drivers for retailers (large, mid size or small) as better hardware performance, increase server capacity and channel capacity and raising IT costs. Therefore, the discussed IT delivery model would be gradual to cater the need of any specific retail through guaranteed performance of applications.